\documentclass[9pt,twocolumn,twoside]{opticajnl}
\journal{opticajournal} 

\setboolean{shortarticle}{false}

\usepackage{lineno}
\usepackage{siunitx}

\title{Low threshold integrated optical parametric oscillator with a compact Bragg resonator}

\author[1,\dag]{Jost Kellner}
\author[1,*,\dag]{Alessandra Sabatti}
\author{Andreas Maeder}
\author{Rachel Grange}

\affil[1]{ETH Zurich, Department of Physics, Institute for Quantum Electronics, Optical Nanomaterial Group, CH-8093 Zurich, Switzerland}
\affil[*]{asabatti@phys.ethz.ch}
\affil[$\dag$]{These authors contributed equally to this work.}

\begin{abstract}
Optical parametric oscillators (OPOs) have been studied as basic components for optical computing with phase encoding and Ising machines. Integrated photonics offers a scalable solution to incorporate a progressively larger number of devices towards a functional computing module. Among the available platforms, lithium niobate on insulator is an excellent candidate for this goal thanks to its large second order nonlinearity, which can be leveraged via periodic poling of the thin film.  In this work, we show a device with a 2.5 mW threshold for parametric oscillation, which is the lowest reported to date among double-resonant OPOs. We use a novel configuration with a Fabry-Pérot cavity, which reduces the footprint compared to a typical ring resonator by a factor 10. 
Tuning our devices using pump wavelength and local heating, we can operate the oscillators at degeneracy, which is crucial for logical operations requiring phase bistability. Our results showcase the device as an ideal building block for phase-encoded integrated optical computing, enabling spatial multiplexing with reduced footprint and power consumption. 
\end{abstract}

\setboolean{displaycopyright}{false} 

\begin{document}

\maketitle

\section{Introduction}
In the last years optical computing has emerged as a potential solution to the increasing demand for fast data processing, also thanks to the maturity of integrated photonic platforms
\cite{OpticalComputing,IntegratedNeuro}. Neuromorphic computational architectures are good candidates for efficiently solving problems that are computationally inaccessible for traditional computing. In particular, Ising machines gained attention due to the correspondence between the ground state of the Ising Hamiltonian and the solution of NP-hard problems \cite{Barahona, MarandiTheory}. One proposed scheme for implementing Ising machines is a network of coupled optical parametric oscillators (OPOs). When operating at degeneracy, OPOs display a phase bistability that can be exploited to mimic the two possible Ising spin orientations \cite{CoherenceOPO}.
The first demonstrations of coherent Ising machines were realized in free space and fiber optics \cite{100_spin_all_to_all,PhaseTransitions,100_000_spin,MarandiFreeSpace}.
More recent experiments have focused towards integrated photonics implementations, to increase scalability and lower the power consumption. 
Even though efforts have been made to realize photonic integrated Ising machines exploiting the $\chi^3$ nonlinearity in the SiN platform with a spatial multiplexing scheme \cite{DOPO}, and using the quadratic $\chi^2$ nonlinearity of lithium niobate in a time-multiplexed configuration \cite{Marandi70}, the spatial coupling of OPOs leveraging the $\chi^2$ nonlinearity has not been investigated.

The building block for realizing such a network with spatial coupling has some crucial requirements, namely a small footprint for dense device integration, a low threshold for enabling the simultaneous operation of multiple OPO units with sustainable on-chip power, and, most importantly, the possibility of operating all the OPOs at degeneracy, oscillating at the same frequency. Widely adopted tuning mechanisms rely on pump wavelength tuning and heating of the substrate. \cite{ultralowPower, UltralowThreshold}.
Quadratic OPOs have been realized in the lithium niobate on insulator (LNOI) platform in a double and triple resonant configuration \cite{Widely_tunable_OPO, ultralowPower, KellerFejer}, using both an x-cut and z-cut lithium niobate thin film \cite{UltralowThreshold}. The advantage of a triple resonant configuration is in general the low threshold due to the resonant pump. Despite this, their fundamental limitation is the overlap of pump and half signal resonance frequencies. This is typically overcome by temperature tuning, yet different devices on the same sample will inevitably require different bias points, due to geometrical imperfections, thus hindering the ultimate goal of coupling different degenerate devices. On the other hand, the double resonant OPO offers more flexibility, by allowing the pump to be set to an arbitrary frequency, but it typically requires larger pump power and large footprint devices. 

\begin{figure*}[htp]
    \centering
    \includegraphics[width=0.75\textwidth]{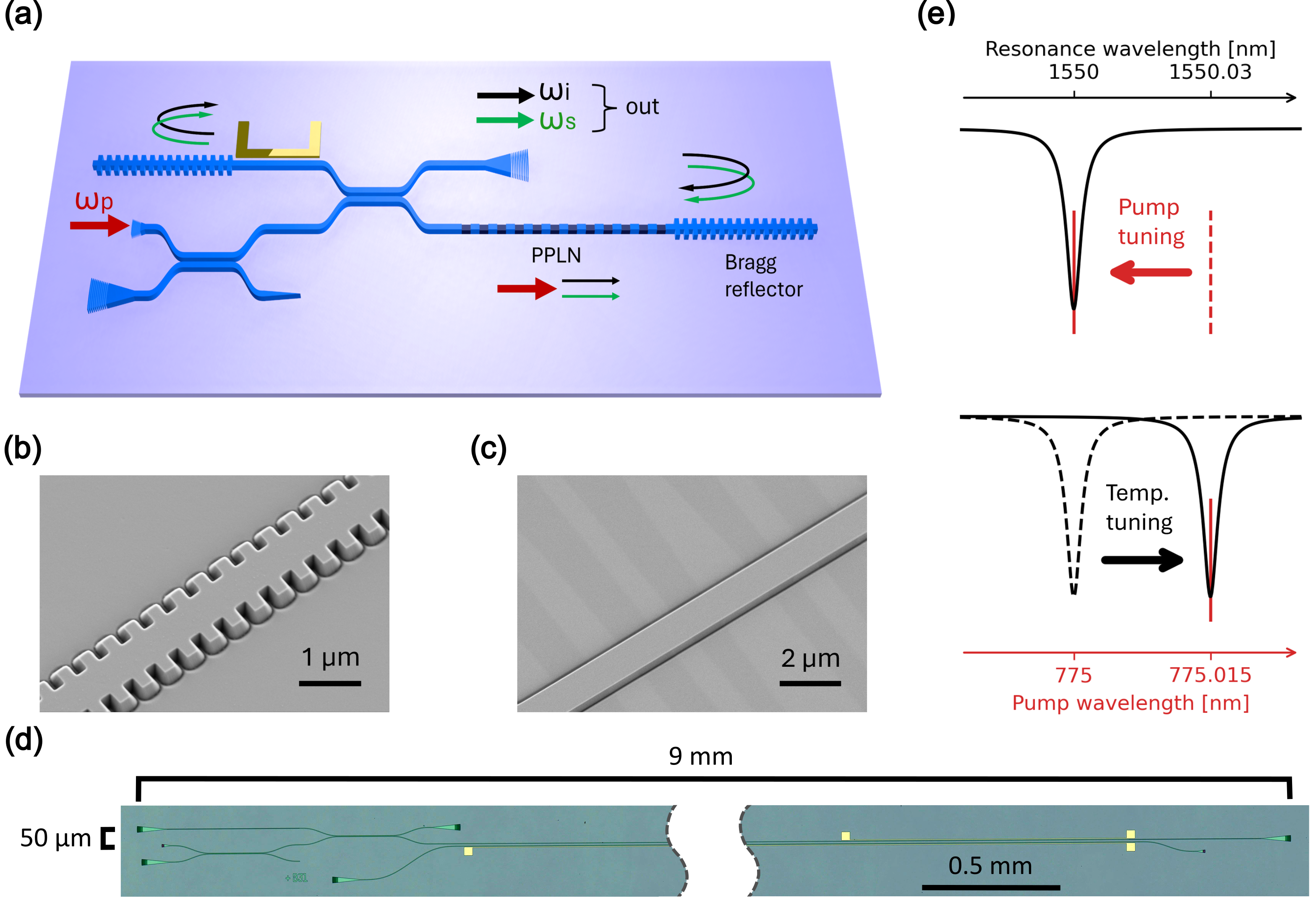}
    \caption{Device concept and overview. (a) Sketch of the device showing the Fabry-Pérot cavity, the PPLN waveguide and the thermo-optic phase shifter. (B) SEM image of a Bragg reflector. (c) SEM image of a PPLN waveguide, with the domain pattern revealed by an intensity contrast in the electronic signal. (d) Mechanisms for tuning the OPO through the pump wavelength (top) and through the thermo-optic shift of the signal resonances (bottom).
    }
    \label{fig:fig1}
\end{figure*}

Here, we achieve a low threshold and small footprint by designing a double resonant OPO in a novel configuration, using a Fabry-Pérot cavity, with two integrated Bragg reflectors (BRs) \cite{GalloBragg, MarcBragg, DavidBragg}. Our approach enables to reduce the device footprint compared to standard ring cavities by at least a factor 10 \cite{Widely_tunable_OPO, squeezingOPO, RFOPO, Marandi70, MidIR_spectro}. Simultaneously, we demonstrate a record low threshold value of \SI{2.5}{mW} for double resonant OPOs. 
Selection of the operating regime, thus controllable switching between degenerate and non-degenerate cases, is implemented with pump laser tuning featuring a sub-pm step size, and local heating induced by thermo-optic phase shifters. The latter approach allows to introduce a refractive index variation only in a restricted region, hence enabling simultaneous independent tuning of a large number of devices within the same sample, without breaking out of degenerate operation.
Achieving the small footprint, low threshold and independent tunability, the presented device fulfills the basic requirements for building networks of spatially coupled OPOs for optical computing.

\section{Results}

\subsection{Design}

A schematic of our device is illustrated in Fig. 1(a). Two Bragg reflectors (BRs), displayed in the SEM image in Figure \ref{fig:fig1}(b), are used to form a Fabry-Pérot cavity that presents a reflection band in the signal and idler wavelength range. Optical parametric amplification is realized in a periodically poled lithium niobate (PPLN) section inside the resonator. Figure \ref{fig:fig1}(c) shows an SEM picture of the nonlinear waveguide, with the neighboring domains characterized by a varying signal intensity. Inside the cavity, a directional coupler is utilised as a wavelength de-multiplexer (WDM), such that the signal is transmitted at the cross port and the pump at the bar port. This configuration is an alternative to injecting the pump directly through the Bragg reflector (BR), as the latter works as a high-order Bragg grating for the pump and deflects it out of plane. As shown in the schematics, the circuit has another WDM at the input, that allows to input the fundamental harmonic through the cross port and linearly characterise the cavity. This part of the circuit is used only for linear characterisation, but it does not add any functionality, thus it does not contribute to the device footprint, in the perspective of integrating coupled devices. The readout is performed through the top-right arm of the WDM inside the cavity for both the OPO operation and the linear measurement. 
A phase-shifter close to the waveguide in the Fabry-Pérot cavity is meant to locally tune the effective cavity length \cite{AndreasTOPS}, hence to tune the spectral position of the resonances as a function of the applied power, without changing the other properties of the device. 
An optical microscope picture of the device reported in Figure \ref{fig:fig1}(d) shows that the Fabry-Pérot configuration allows to reduce the OPO size to \SI{50}{\micro\metre} in the direction orthogonal to the waveguides.
To reach the degenerate mode, due to energy conservation, the pump frequency has to be exactly twice the resonant frequency that will be excited for both signal and idler. With this device, we can achieve it with two independent tuning mechanisms, as represented in Figure \ref{fig:fig1}(e). The first one is the fine-tuning of the pump wavelength, and the second one consists of shifting the resonance position to match the frequency of the pump laser using the thermo-optic phase shifter. The implementation of independent methods for the tuning allows to operate different devices at degeneracy and at the same frequency, provided by a unique laser source.  

\subsection{Measurements}

\begin{figure}
    \centering
    \includegraphics[width=1\linewidth]{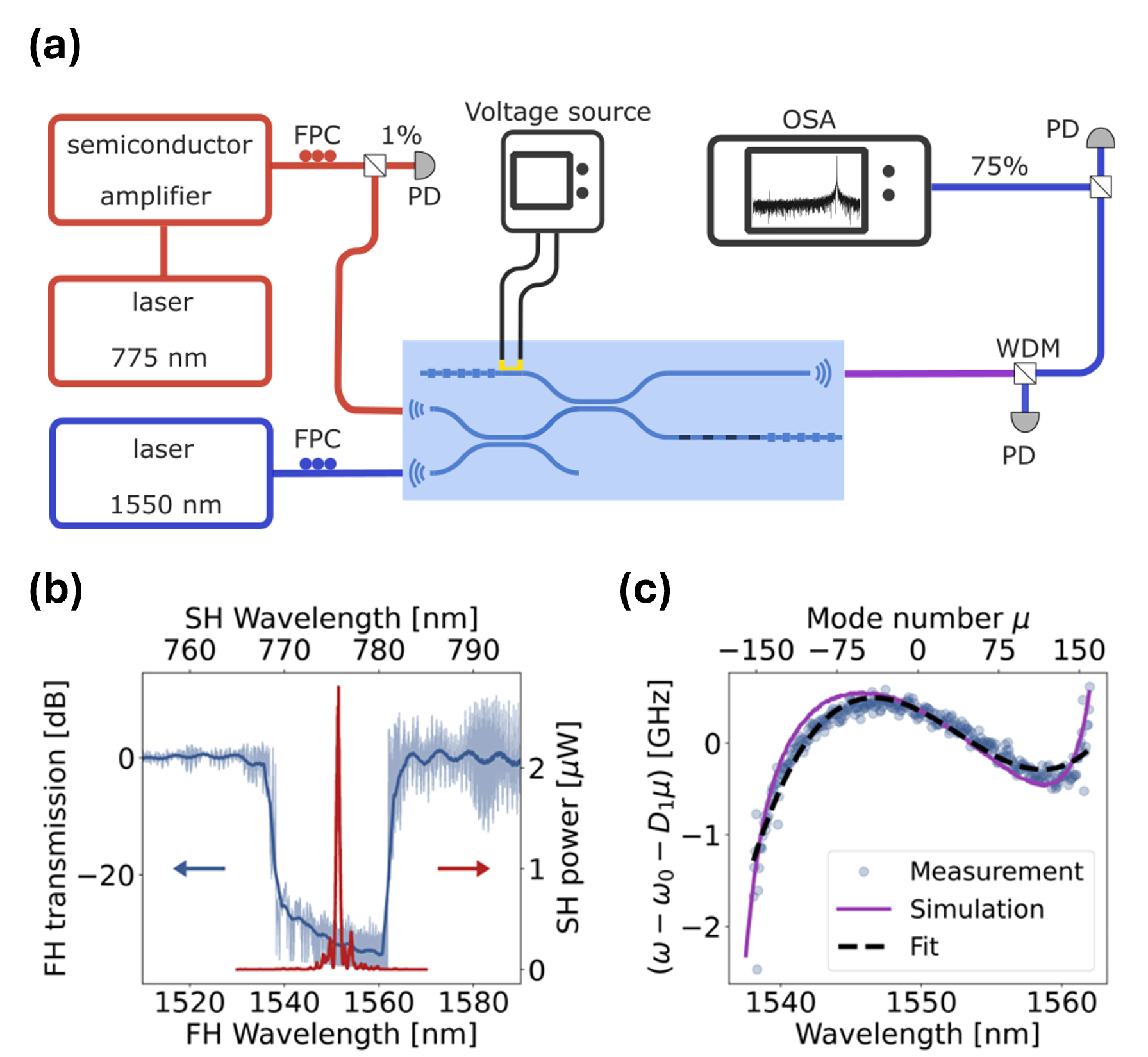}
    \caption{Characterisation of nonlinear devices. (a) Experimental setup. See the text for more details. (b) Measured Bragg grating stop band (blue curve) and second-harmonic signal measured from a calibration waveguide in the same poling region. (c) Measured and simulated dispersion. FPC fiber polarization controller, PD photodiode, OSA optical spectrum analyzer, WDM wavelength demultiplexer.}
    \label{fig:fig2}
\end{figure}

In the measurement setup sketched in Figure \ref{fig:fig2}(a), the elements drawn in blue represent light at \SI{775}{nm}, while the red ones at \SI{1550}{nm} (see Methods section). 

First, we report the characterisation of the OPO building blocks. The stop-band of the BRs is probed by injecting light at the fundamental harmonic (FH) through a grating coupler connected to the BR and measuring the transmission at the readout grating. The stop-band spectrum is plotted in Figure \ref{fig:fig2}(b) in blue. It features a \SI{25}{nm} bandwidth and an extinction ratio that varies between \SI{25}{dB} and \SI{33}{dB}. This spectrum is not the pure transmission of the BR on the left, but it is rather a combination of the stop-band and the cavity transmission. Therefore we observe the presence of Fabry-Pérot resonances inside the stop-band. Since the signal inside the stop-band approaches the noise floor of the detector, the resonance characterisation was performed by injecting light from the input grating at the FH. The measured spectrum presents resonances with Q-factor around $5 \cdot 10^{5}$ as reported in the supplementary material. Figure \ref{fig:fig2}(c) displays the dispersion $D_2$ of the cavity resonant frequencies according to the expression $D_2(\mu) = \omega(\mu) - \omega_0 - \mu D_1 $, where $\mu$ is the mode number, $\omega(\mu)$ is the frequency of the mode $\mu$, $\omega_0$ is the center frequency of the fit and $D_1$ the free spectral range, which is \SI{70}{pm} for the considered cavity. The dispersion curve follows a trend that is relatable to the phase given to the FH by the reflection at the BR. The experimental data in Figure \ref{fig:fig2}(c) is compared with the simulated dispersion of the resonances of a Bragg cavity using the transfer matrix method, confirming a consistent trend. The measured dispersion is fitted with a polynomial function, which will be used in the next section for simulating the tuning of the OPO in the ideal case of a device with a smooth dispersion function.

The quasi-phase matching position is obtained through a second-harmonic measurement performed on a twin waveguide located in the same periodically poled region, but independent from the OPO, in which the poled domains are \SI{30}{\micro\metre} wide, allowing to accommodate two non-coupled waveguides. Figure \ref{fig:fig2}(b) shows the second harmonic signal as a function of the fundamental input wavelength in red, which overlaps with the center of the stop-band. This ensures that both signal and idler are supported by the cavity with the effective bandwidth of the double resonant OPO coinciding with the stop-bands of the BRs. The measured estimated SH harmonic efficiency is \SI{1155}{\percent/W cm^2}.


\begin{figure*}
    \centering
    \includegraphics[width=0.9\linewidth]{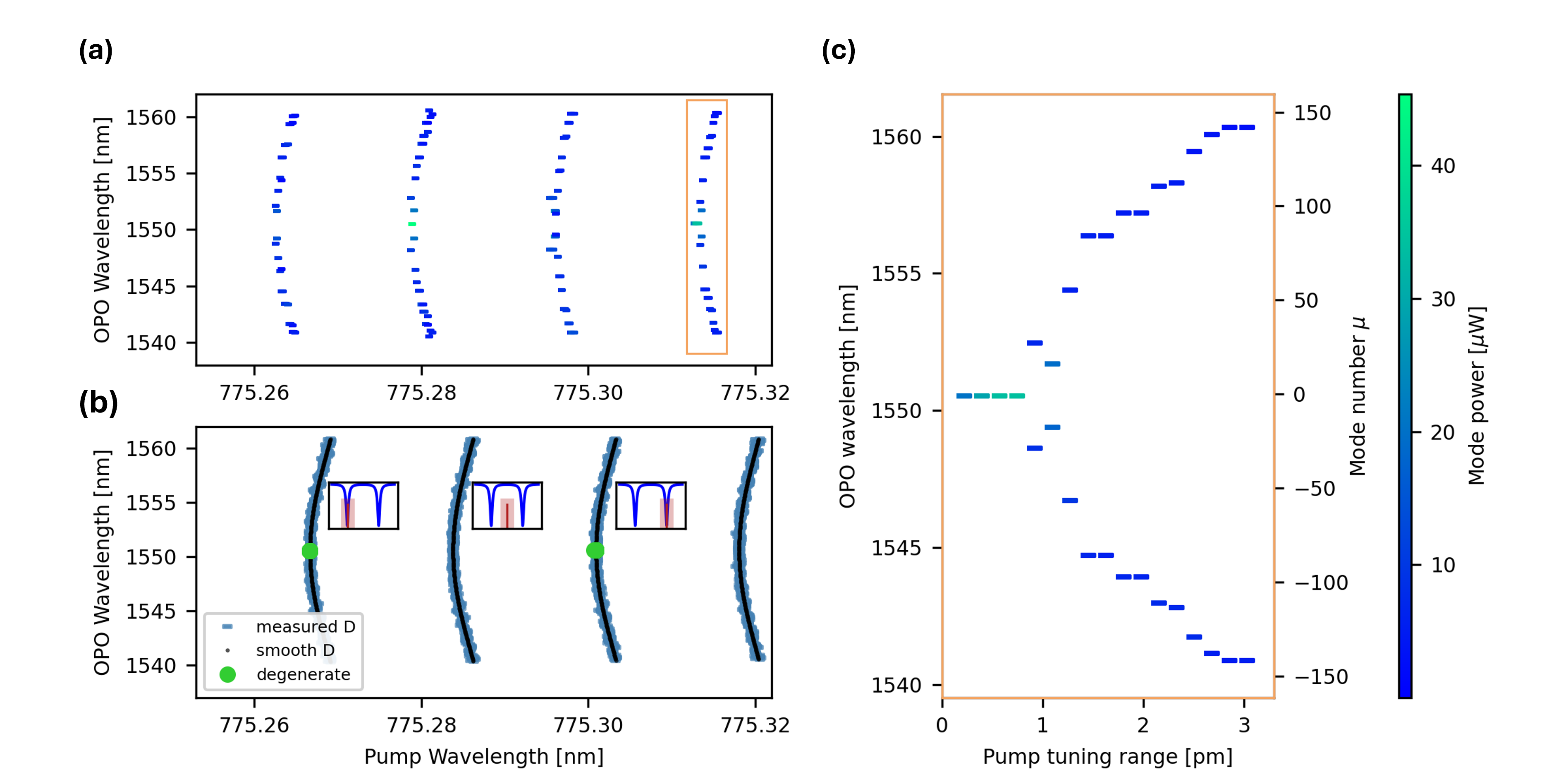}
    \caption{Tuning between degenerate and non-degenerate optical parametric oscillation through pump wavelength tuning. (a) Measured signal and idler wavelength as a function of pump detuning. The tuning curve has a periodicity of one quarter of the resonator free spectral range and presents the degenerate mode once every two periods. (b) Simulation of the tuning behaviour based on energy conservation for the measured dispersion (light blue) and for the dispersion fitted in Figure \ref{fig:fig2} in black. The insets show the relative position between the pump and the signal resonance correspondingly to tuning curves with or without the degenerate mode. (c) Zoomed in tuning plot with OPO wavelength and corresponding mode number, showing the tunability into the degenerate and non-degenerate regime.}
    \label{fig:3}
\end{figure*}

To study the frequency response of the integrated OPOs, 
we tune the pump laser from \SI{775.25}{nm} to \SI{775.32}{nm} and display its spectrum in Figure \ref{fig:3}(a). The tuning spans from spectra with signal and idler close to degeneracy to spectra up to \SI{20}{nm} separation. The plot shows that the supported modes are excited only for some subsets of the pump range. These subsets have a periodicity of half the free spectral range of the cavity fundamental modes. From the plot, it is possible to see that the degenerate mode is excited only once every two subsets. This behavior can be explained by the relative spectral position of the pump with respect to the signal resonances of the cavity, and it is further investigated and reproduced in the simulation (Figure \ref{fig:3}(b)). This plot shows the simulation of the pump tuning by taking into account only energy conservation for the measured frequency of the resonances. The relative shift in wavelength between the experimental and simulated plots is due to variations in the temperature at which the linear spectrum and the OPO spectra were measured.

As represented in the insets of Figure \ref{fig:3}(b), in the left-most supported subset, the half frequency of the pump laser is exactly on top of the resonance at the fundamental, while in the second one, it is in between the two resonances, and this trend repeats periodically in the next subsets. Crucially, only in the first case, it is possible to achieve degenerate operation for the OPO, which is represented with green dots in the simulation. In the second case, because of energy conservation between pump, signal and idler, the operation mode will always be non-degenerate, and the pair nearest to degeneracy will be formed by the two resonances closer to the half pump. Note that the stop-band width limits the bandwidth of the signal and idler in the subsets, but could in principle extend more with a BR with a wider reflection band. 

To study the OPO tunability in more detail, we zoom into one of the subsets of supported modes and report a tunability from degenerate to \SI{20}{nm} separation by sweeping the pump laser by only \SI{3}{pm}. The device is tunable in discrete steps because only signal and idler pairs that are both resonant in the cavity can reach the threshold. Due to the chromatic dispersion of the modes (Figure \ref{fig:fig2}(c)), when the pump is detuned, a different mode pair matches the pump energy and gets excited. Without dispersion in the resonance frequencies, the tuning would not be possible, as all the modes would equally fulfil the energy conservation only at the very discrete cases of the half pump being exactly on top of a resonance, or exactly between two resonances. 
The tuning curve is non-monotonic, as it inherits the disorder of the mode dispersion, shown in Figure \ref{fig:fig2}(c). In fact the mode number does not always increase monotonically with the pump wavelength. In the simulation in figure \ref{fig:3}(b) it is shown how the mode selection is partially disordered when considering the experimental mode distribution. Using the resonances with a dispersion fitted as in Figure \ref{fig:fig2}(c), the tuning function, represented in black, is monotonic. Another feature in the experimental tuning curve is the presence of only a subgroup of all the mode numbers. This can be explained by the competition between different signal idler pairs, that present different Q-factors (See supplementary). In fact, close to the threshold, only the pair with the best loss to parametric gain relation will be excited and measured in the spectrum.

\begin{figure}
    \centering
    \includegraphics[width=1\linewidth]{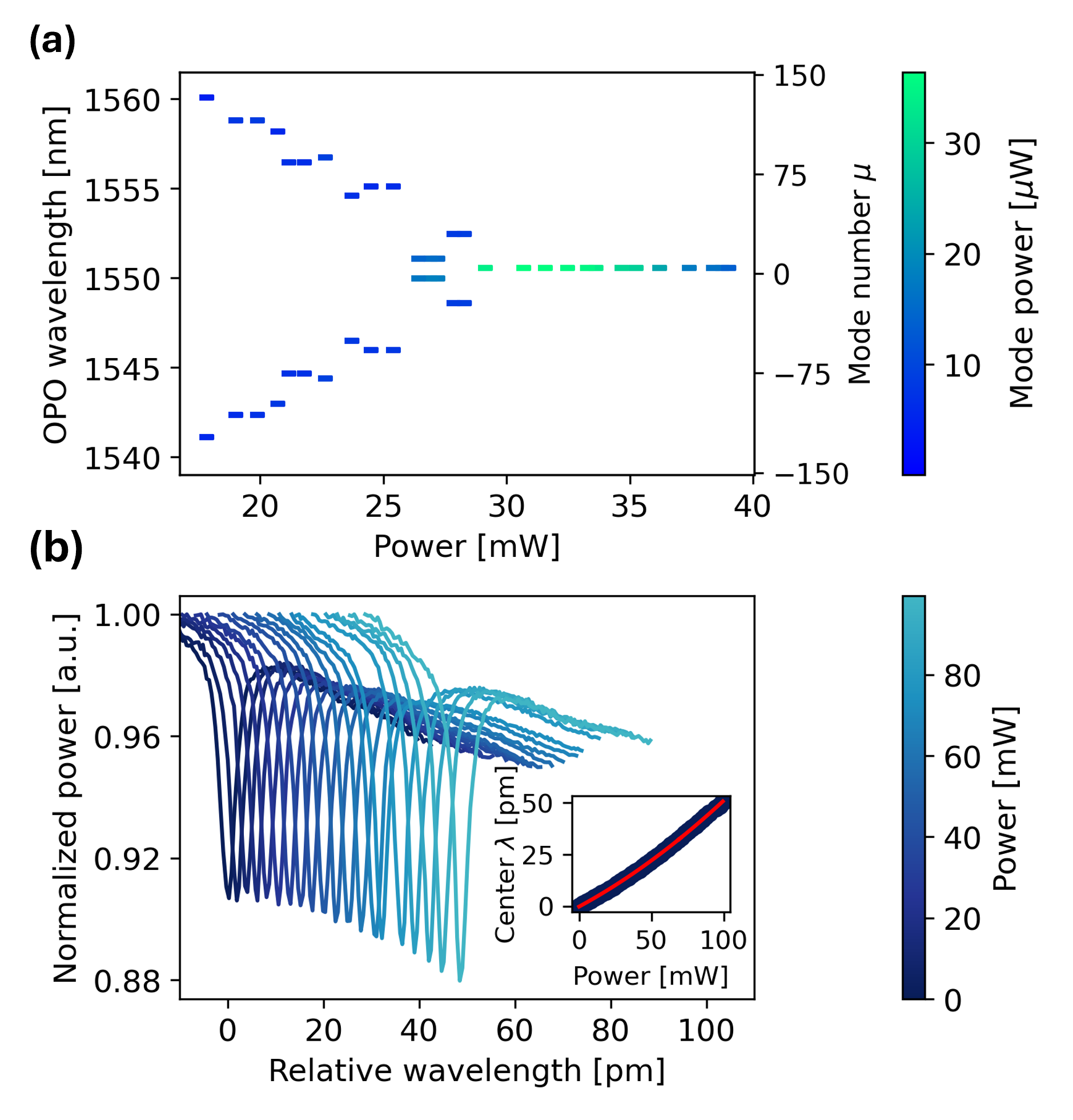}
    \caption{Tuning between degenerate and non-degenerate optical parametric oscillation through sweeping the power applied to the thermo-optic electrode. (a) OPO spectra as a function of power applied to the electrode. (b) Resonance plot for different thermo-optic powers and resonance frequency shift as a function of thermo-optic power (inset). }
    \label{fig:Temperature}
\end{figure}

The selection of different signal and idler mode pairs can also be performed through the thermo-optic tuning of the cavity, by sweeping the electric power applied to the electrode, for a a pump wavelength of \SI{775.38}{nm}, as shown in Figure \ref{fig:Temperature}(a). The power dissipated in the metal electrode locally changes the temperature of a portion of the waveguide in the resonator, resulting in a refractive index change and thus an effective change in the cavity length. This change is measured in Figure \ref{fig:Temperature}(b), which reports the position of the resonances as a function of the applied power. The inset shows the trend of the center position of the resonance as a function of applied power, which is linear in first approximation \cite{TOPSnonlinear}. Since, when sweeping the power, the resonant peak is red-shifted with respect to the pump, unlike for the previous tuning mechanism, the tuning curve displays the opposite orientation with respect to the x-axis.

\begin{figure}
    \centering
    \includegraphics[width=1\linewidth]{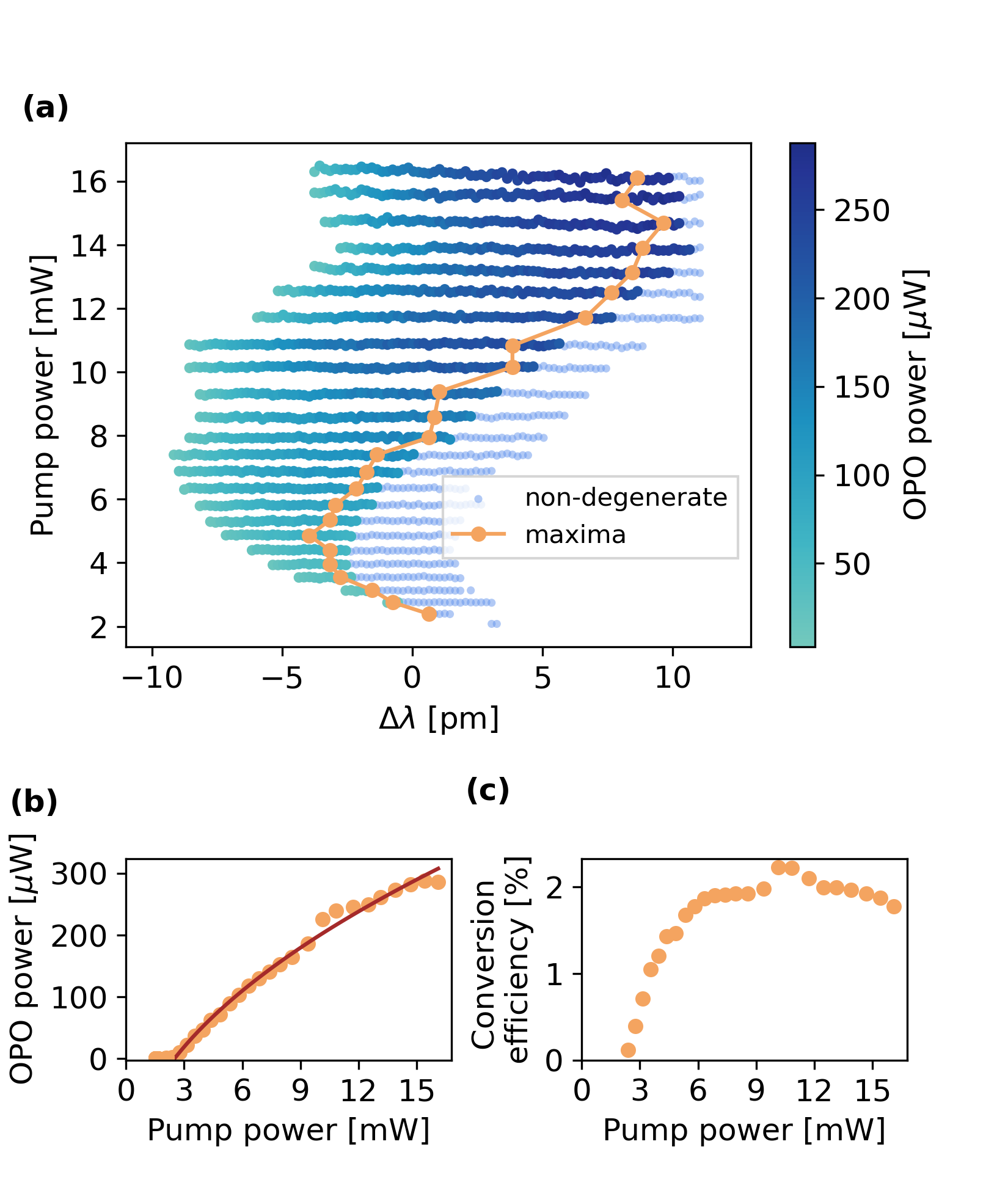}
    \caption{Threshold study for the OPO in the degenerate mode. (a) Pump wavelength sweeps for increasing pump power. The maximum power of each sweep is indicated in orange and the non-degenerate OPO excitations are plotted in light blue. (b) OPO output power as a function of pump power exhibiting the threshold behavior and the characteristics square root trend. (c) Conversion efficiency of the OPO.}
    \label{fig:threshold}
\end{figure}

The OPO threshold is characterised specifically for the degenerate mode, unlike the typical threshold measurement method, which consists in sweeping the pump power and measuring the output power, regardless of which mode is excited. As represented in Figure \ref{fig:threshold} (a) we perform a series of pump wavelength sweeps with increasing pump power. We observe that the region of stability of the degenerate mode becomes larger for increasing pump power. We track the maximum OPO power for each dataset corresponding to a different pump power and represent it with the dots in Figure \ref{fig:threshold}(a). We expect the OPO power to be larger in the middle of this stability region, but for positive detuning the other modes come into play and the spectrum is either non-degenerate or a combination of degenerate and non-degenerate, and in this last case the OPO power gets redistributed between the different modes. Therefore, the maximum OPO power is measured close to the transition towards other modes. For each dataset with the same pump power, we collect the points with higher output power and use them to construct the threshold plot of the degenerate mode in (\ref{fig:threshold}(b)). We measure a threshold power of \SI{2.5}{mW}, which is lower than other examples of double resonant OPOs reported in literature \cite{Widely_tunable_OPO, RFOPO,  MidIR_spectro, squeezingOPO, Marandi70}. The conversion efficiency for the same measurement is shown in panel (c), reaching a maximum conversion efficiency above \SI{2}{\percent}. The reported efficiency is limited by the reduced escape efficiency of the signal from the cavity, meaning that the low threshold comes at the expense of the intensity of the output signal. However, this is not a problem for the intended application, as the measurement of the OPO phase state does not require a large power generation.

\subsection{Discussion}

This work presents an integrated device that meets the criteria for establishing a network of coupled optical parametric oscillators (OPOs). The device features a compact footprint of \SI{0.45}{mm^2}, a low threshold power of \SI{2.5}{mW}, and a mechanism for tuning multiple OPOs to operate at degeneracy at the same frequency. The comparison with the literature is shown in Table \ref{tab}.
We fulfilled these criteria by designing the cavity in a Fabry-Perot configuration, which significantly reduces the area of the device. The threshold was minimized by lowering the power coupled out of the cavity and engineering the efficiency of the PPLN for the optical parametric amplification. 
Regarding the last criterion, we studied the tuning spectra and determined the conditions for reaching the degenerate mode.
Because even resonators with the same design present resonances that are characterized by an overall spectral offset, due to tiny variations in the different waveguide devices, coupling two or more OPOs and setting them all independently in the degenerate mode for the same pump frequency is challenging.
The problem can be solved with this device by tuning the pump frequency to excite the degenerate mode for the first OPO and temperature-tuning the resonance position of all the others independently. The non-resonant pump is crucial as it facilitates access to the complete tuning curve, including the degenerate mode.

Additionally, we also characterized the same device type in a single resonant configuration, for which the signal is resonant inside the stopband, with a non-degenerate and non-resonant idler. In principle, the idler could be smoothly tuned in the infrared region by sweeping the pump laser (more details in the supplementary).

In conclusion, we showed that reducing the footprint of integrated OPOs while simultaneously obtaining low oscillation threshold and independent tunability is possible with our novel device configuration. This enables the realization of OPO networks which can implement phase-encoded optical computing in a spatially multiplexed configuration. Our results demonstrate that such operation is not only possible, but is also power efficient from both an optical and electrical perspective, and paves the way to alternative and innovative computing schemes.
\begin{table}[htp]
    \centering
    \begin{tabular}{c|c|c|c|c}
        Reference & Threshold & Footprint & Conversion \\
        & [\SI{}{mW}] & [\SI{}{mm^2}] & efficiency [\SI{}{\%}] & \\
        \hline
         \cite{Widely_tunable_OPO} & \SI{32}{^*} & \SI{4.2}{} & $15 $ \\
         \cite{squeezingOPO} & \SI{25}{} & $\sim$\SI{8.4}{} & $<1$  \\
         \cite{RFOPO} & \SI{80}{} & \SI{10}{} & $34$ \\
         \cite{MidIR_spectro} & \SI{80}{} & \SI{9.6}{} & $15 $ \\
         \cite{OPO_comb} & \SI{50}{^*} & - & $5$ \\
         This work & \SI{2.5}{} & \SI{0.45}{} & $2$ \\
    \end{tabular}
    \caption{Comparison of threshold and footprint between this work and other double resonant OPOs. $^*$ Peak power in pulsed regime.}
    \label{tab}
\end{table}

\subsection{Methods}

\textbf{Fabrication}
The circuit is realized using the LNOI platform, with a \SI{5}{\%} MgO doped LN film whose thickness is \SI{300}{nm}, and with a \SI{2}{\um} thick buried oxide. The periodic poling is realized prior to waveguide etching, using \SI{100}{nm} thick Cr electrodes for the high voltage pulse application. The electrodes are patterned by electron-beam lithography (EBL), electron-beam evaporation and liftoff. A photoresist that serves as an electrical insulation layer is applied to the chip during the poling process. After the removal of the photoresist and of the poling electrodes, the waveguides are patterned into a FOX16 mask with another EBL step. The film is etched by \SI{200}{nm} using an Argon ion etching process via ICP-RIE \cite{FabPaper}. The LN redeposited during the etching is removed with an RCA-SC1 cleaning step, followed by a BHF dip to remove the remaining mask. The phase shifters are realized with a third EBL exposure, followed by evaporation of a \SI{100}{nm} thick Au layer with a \SI{5}{nm} Cr adhesion layer and liftoff. 
To exploit the full bandwidth defined by the stop-band, the SH peak had to be aligned with the center of the stop-band with a few nanometers precision. The fabricated device presented at first a red mismatch of about \SI{20}{nm} of the SHG with respect to the stop-band. This was compensated for by adding an atomic layer deposited film of about \SI{11}{nm} of silicon dioxide on the whole chip \cite{LoncarWaferScale}, which changes the effective refractive indices of the FH and second harmonic. The phase matching spectrum was blue-shifted by \SI{20}{nm} and the stop-band was red-shifted by \SI{2}{nm}, ensuring a good overlap.

\textbf{Measurement setup}
Linear characterisation of the OPO cavity and nonlinear frequency conversion is performed using a tunable laser at 1550 nm. We couple the laser to the chip and split the output using an off-chip WDM. Both powers are detected on separate InGaAs photodetectors.  
To probe the OPO behavior, a tunable laser source is amplified via a tapered semiconductor amplifier. We first couple the pump to the chip with a value below the threshold. By adjusting the pump wavelength and measuring the intensity of the OPO below the threshold, we find the strongest response. The pump laser allows coarse tuning in \SI{10}{pm} steps and the control of the intracavity piezo for finer tuning. Polarization is controlled by a fiber-coupled polarization controller. The input power is constantly monitored by measuring \SI{1}{\percent} with a photodiode before coupling. The OPO output is collected and the OPO signal and idler are split from the possible remaining pump light using an off-chip WDM. Output power and spectrum are detected simultaneously on a power meter and an optical spectrum analyzer respectively. Coupling to and from the chip is performed with cleaved single-mode fibers and grating couplers with a specific design for the pump at \SI{775}{nm} and the signal and idler at \SI{1550}{nm}.

\begin{backmatter}

\bmsection{Funding}
This work was supported by the Swiss National Science Foundation Grant (project number 206008).

\bmsection{Acknowledgment}
We acknowledge support for the fabrication and the characterisation of our samples from the Scientific Center of Optical and Electron Microscopy ScopeM and from the cleanroom facilities BRNC and FIRST of ETH Zurich and IBM Ruschlikon. We thank Giovanni Finco and Pierre Didier for contributing to revise the manuscript.

\bmsection{Disclosures}
The authors declare no conflicts of interest.

\bmsection{Data availability} Data underlying the results presented in this paper are not publicly available at this time but may be obtained from the authors upon reasonable request.

\bmsection{Supplemental document}
See Supplement 1 for supporting content.

\bmsection{Author contributions}
J.K. and A.S contributed equally to this work.

\end{backmatter}


\bibliography{bibliography.bib}

\end{document}